Title:

# Experimental Evidence for Using a TTM Stages of Change Model in Boosting Progress Toward 2FA Adoption

Authors:

Cori Faklaris, Laura Dabbish and Jason I. Hong, *Carnegie Mellon University*

Corresponding author:

Cori Faklaris, cfaklari@cs.cmu.edu, Human-Computer Interaction Institute, Carnegie Mellon University, 5000 Forbes Ave., Pittsburgh, PA 15213 (Permanent email: cori@corifaklaris.com)

Funding:

This work was supported by the U.S. National Science Foundation, grant no. CNS-1704087. The first author also received fellowship support from the CyLab Security and Privacy Institute and the Center for Informed Democracy and Social Cybersecurity, both at Carnegie Mellon University. Sponsors were not involved in any phase of research or article preparation.

# Abstract

Behavior change ideas from health psychology can also help boost end users' compliance with security recommendations, such as adopting two-factor authentication (2FA). Our research adapts the Transtheoretical Model's Stages of Change from health and wellness research to a cybersecurity context. We first create and validate an assessment to identify workers on Amazon Mechanical Turk who have not enabled 2FA for their accounts as being in Stage 1 (no intention to adopt 2FA) or Stages 2-3 (some intention to adopt 2FA). We randomly assigned participants to receive an informational intervention with varied content (highlighting process, norms, or both) or not. After three days, we again surveyed workers for Stage of Amazon 2FA adoption. We found that those in the intervention group showed more progress toward action/maintenance (Stages 4-5) than those in the control group, and those who received content highlighting the process of enabling 2FA were significantly more likely to progress toward 2FA adoption. Our work contributes support for applying a Stages of Change Model in usable security.

# 1. Introduction

Recent data has painted a despairing picture of people's adoption of expert-recommended security measures such as two-factor authentication (2FA). A Google engineer revealed at a 2018 security conference that fewer than 10 percent of Gmail users had enabled 2FA, in which users first enter a password (Factor 1) and then use a mobile code or security key (Factor 2) to finish logging into their accounts [54]. This number echoes a finding by Pew Research Center that only 10 percent of surveyed Americans could correctly identify a screenshot as depicting a 2FA account login [64]. Without broader adoption by end users, websites' implementation of 2FA may continue to stagnate despite widespread data breaches and password thefts [9] that expose accounts not secured with 2FA to easy attack.

This data underlines the significance and broader impact that could result from adapting proven ideas from other fields such as health psychology for boosting user adoption of 2FA and other security measures. To this end, we apply Prochaska and DiClemente's *Transtheoretical Model of Behavior Change* [44,63]. This model, and its associated *Stages of Change*, has already been identified as a useful framework for privacy and security research [8,10,25,30,55]. It posits that "behavior change is a process that unfolds over time through a sequence of stages" and that individuals need planned interventions matched to their stages of change in order to move them toward desired actions and maintenance of their new behaviors [44]. Similarly, for cybersecurity, we see promise in structuring *which* interventions are given to end users and *when* to most effectively spur their progress toward the adoption and maintenance of expert-recommended security behaviors.

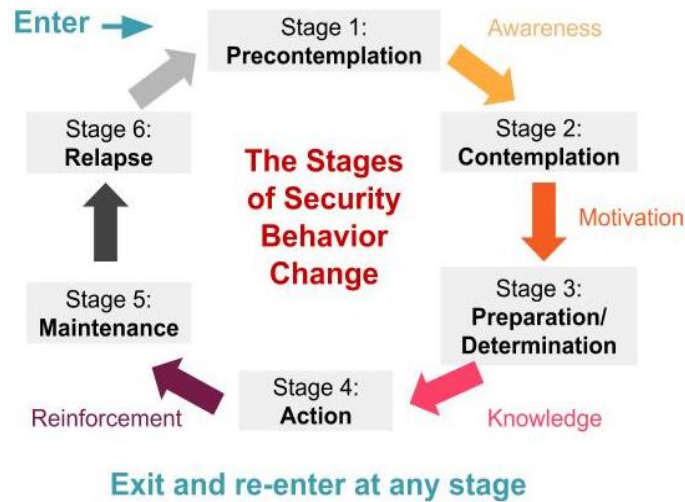

*Figure 1: As drawn from Transtheoretical Model research, the Stages of Change are Pre-contemplation (Stage 1, no intention to adopt the behavior), Contemplation (Stage 2, some intention to adopt the behavior, with a bias for procrastination), Preparation/Determination (Stage 3, some intention to adopt the behavior, with a bias for action), Action (Stage 4, adoption of the behavior within the past 30 days), Maintenance (Stage 5, sustained behavior adoption for two to six months) and Relapse (Stage 6, stopping the behavior, leading back to Stage 1 or other stages). The processes that move people from one stage to the next, such as Awareness (Stage 1 to Stage 2), are marked next to the arrows.*

Drawing on the above literature, we identified two research questions for exploration in this study:

- **RQ1**: *How can we adapt and validate the use of a staging algorithm from health psychology to measure an individual's Stage of Security Behavior Change?*
- **RQ2**: *How do people in different security behavior change stages react to an informational intervention?*

To answer these questions, we adapt prior TTM staging measures to a cybersecurity context and validate our measure against two existing psychometric scales. We then test whether providing 2FA information, as our awareness-oriented intervention, helps to boost end users' progress toward adopting 2FA. Using blind recruitment for a qualification test on the crowdsourcing platform Amazon Mechanical Turk, we identify workers who have not enabled Amazon 2FA as being in Stage 1 (no intention to adopt 2FA) or Stages 2-3 (some intention to adopt 2FA). We randomly assign the workers in Stages 1-3 to view and rate one-page informational handouts: two of which emphasize the process of enabling 2FA, two of which emphasize social norms of 2FA adoption, or one with no 2FA information, as a control. Within a week, we again survey workers for their Stage of Change of Amazon 2FA adoption. We find a statistically significant improvement in progress toward 2FA adoption for those who viewed 2FA informational handouts and started in Stage 1, compared with those who started in Stages 2-3. This suggests that awareness-oriented interventions such as handouts are a match for Stage 1 but not Stages 2-3, as per the Stages of Security Behavior Change Model.

## 1.1. Information and Security Behavior Change

Human information processing [52,53] is a key framework for explaining how perception, cognition and action interact. In this view, the mind itself is like a computer, taking in input stimuli such as typed messages or visual diagrams and encoding and transforming them in memory and cognition to produce a behavior as the output. Prior work that explored this dynamic includes Rogers' *Diffusion of Innovations* theory [48] of how messages spread in a network about a "new ideal," Davis and collaborators' models [21,22] of how user perceptions of usefulness and ease of use predict *Technology Acceptance*, and Bravo-Lillo et al. [7] and Egelman et al. [26]'s applications of the *Communication-Human Information Processing* (C-HIP) model [62] to end-user security warnings. The latter paper notes that users trust information that is well-designed, which phishing and other social cybersecurity attacks exploit, but also helps in designing effective prompts for user actions.

### 1.1.1 Information Security Awareness

Bravo-Lillo and Egelman [7,26] test information delivered as on-screen warnings, in which people's attention must be captured and a sense of danger conveyed to immediately produce a protective behavior. They are not directly concerned, however, with security information that is important but not urgent in the moment, such as educating an app's users about the existence and ease of enabling a security method such as two-factor authentication. Abawajy [1] reviewed the effectiveness of diverse methods of information security awareness falling into three categories: text-based (such as leaflets, posters, newsletters, and websites), game-based (such as CyberCIEGE [37] and Anti-Phishing Phil [51]) and video-based (such as online training videos and simulators). A combination was found to be more effective than any single one. More recent work by Albayram et al. [6] tested video tutorials about 2FA and assessed the qualities that were associated with users' willingness to try 2FA and to enable 2FA. Participants' positive ratings of the risk and self-efficacy videos for interestingness, informativeness and usefulness were significantly correlated with decisions to enable 2FA. Our work will test whether using a combination of text-based information about 2FA is as effective as focusing on the process of how to enable 2FA, as measured with the Stages of Security Behavior Change.

## 1.1.2 Security Sensitivity

*Security sensitivity* is key to understanding influences on cybersecurity acceptance and adoption. Das defines this concept as "the awareness of, motivation to use, and knowledge of how to use security tools" [15]. His empirical research documented how lack of information reduces sensitivity due to people not correctly perceiving their danger of falling victim to a security breach and failing to register the existence of tools to protect them and their close ties against such threats [15,17–19]. Work that follows on this insight uses Ajzen's *Theory of Reasoned Action* and *Theory of Planned Behavior* [2–5,5], Cialdini's *Weapons of Influence* [11–13,36] and Fogg's *Behavior Change Model* [32–35] to document the influence of observable factual and social information [19,28,31], such as exposure to Facebook's Trusted Contacts tool [18] and media coverage of security breaches [20], in prompting end users' security behaviors [16]. Our work will test whether drawing attention to social norms of 2FA use is as effective as focusing on the process of how to enable 2FA, as measured with the Stages of Security Behavior Change.

## 1.2. Transtheoretical Model of Security Behavior Change

In order to better model how people make security decisions over time, Faklaris [29,30], Ting [55] and others [8,10,25,55] have adapted a framework from health psychology, Prochaska and DiClemente's *Transtheoretical Model of Behavior Change* [44,63]. The TTM marks a shift from thinking of behavior change as occurring in a single, decisive moment to that of a longer-term, cyclical process in which people balance pros and cons along with self-efficacy and temptation in their decision making. This model (Figure 1) integrates leading theories of psychotherapy and behavior change [44]. It posits six Stages of Change: *pre-contemplation* (Stage 1), *contemplation* (Stage 2), *preparation/determination* (Stage 3), *action* (Stage 4), *maintenance* (Stage 5) and *relapse* (Stage 6, leading back to Stage 1 or other stages), together with associated Processes of Change, that define this cyclical and temporal process.

### 1.2.1 Information and the Stages of Change Model

A key assumption that TTM shares with the theories of Reasoned Action and Planned Behavior and the Fogg Behavior Model is that people are not inherently motivated to act without an external prompt or nudge, and that they need help in gaining the necessary ability or affordance for effective action. The TTM Stages of Change are a way to better match a prompt or helpful tool with the individual's stage of willingness to adopt a new behavior. Stage 1 individuals, for instance, may be uninformed about the consequences of their lack of action, and are more likely to respond to *awareness-oriented* informational campaigns than promotions [44]. However, both Stage 2 and Stage 3 individuals are aware of and can recall the pros of changing their behavior, though those in Stage 2 dwell on the cons without sufficient motivation to overcome their procrastination. Stage 3 are more likely to respond to *action-oriented* interventions such as promotions that help them to act on knowledge of why behavior change is desirable [44].

### 1.2.2 Applications of the Stages of Change Model

In medicine and public health, the TTM has been used to tailor messaging and other interventions to move people toward exercise [38], smoking cessation [24,57] and sobriety [41]. Noar et al. found in a meta-analysis of 57 studies using print communications for health behavior change [42] that the type of material used and the use of TTM constructs such as the Stages of Change were associated with significantly greater effect sizes for print communications tailored for individuals, while tailoring on non-TTM constructs such as social norms did not produce significant gains.

In security and privacy, Sano et al. [49,50], Faklaris et al. [30], and Ting et al. [55] have explored applying the Stages of Change and Processes of Change to end user studies. These researchers identified a theoretical and/or empirical basis for classifying computer users by whether they are in either precontemplation (Stage 1), contemplation/preparation (Stages 2-3), or action/maintenance (Stages 4-5) of adopting practices such as updating their operating systems, checking for https in URLs, and using antivirus software. Sano et al. [49,50] tested messaging strategies by stage, for example, finding that a message emphasizing ease of the OS update was significantly associated with users in the preparation stage answering "I update OS now" to a survey item.

### 1.2.3 Evaluating the Stages of Change Model

The TTM and other stage models are not without their critics. Some such as Weinstein et al. [59,60] have challenged tests of stage theories that rely on cross-sectional research designs as not persuasive of their effectiveness for behavior change. They advocate the use of experiments that include a control and that test for not just whether a stage-matched intervention is effective (such as an awareness intervention for Stage 1) but whether a stage-mismatched intervention is ineffective (such as an awareness intervention for Stages 2-3). Prochaska et al. [45] have laid out a hierarchy of stage-theory evaluation criteria, including *clarity*, *consistency*, *parsimony*, *testable*, *empirical adequacy*, *productivity*, *utility*, and *practicality*. Our study follows [60] in using a within-subjects experimental design and [45] in evaluating the Stages of Security Behavior Change Model.

## 1.3. Hypotheses

Based on the previous work, we propose to test the following: that our Stages of Change measure for security will show sufficient variance to be useful in categorizing end users, that a sufficient number of participants will understand the questions and response logic, and that the measure will associate with scores on existing scales.

- ***H1a:*** *Stronger intention to adopt a security behavior is positively associated with a higher security behavior change score.*
- ***H1b:*** *Security attitude is positively associated with a higher security behavior change score.*

We hypothesize that incorporating social norms in an awareness intervention will be as effective as emphasizing the process of security adoption:

- ***H2:*** *Exposure to information emphasizing either social norms or process of 2FA use will increase the likelihood of progressing toward intention to adopt.*

We hypothesize that a combination of both social norms and process information will be more effective than a single handout that emphasizes social norms or process:

- **H3:** *Exposure to information emphasizing both social norms and process of 2FA use will increase the likelihood of progressing toward intention to adopt.*

We also hypothesize that the response to informational interventions will depend on the person's current progress towards adoption based on the Stages of Change Model (Figure 1) [60]:

- **H4a:** *For people in precontemplation (Stage 1), exposure to information about 2FA will increase progress toward intention to adopt (measured as progress toward Stages 4-5).*
- **H4b:** *For people in contemplation or preparation (Stages 2 or 3), there will be no influence of exposure to a 2FA handout.*
- **H4c:** *For people in precontemplation (Stage 1), exposure to a combination of 2FA information will increase progress toward intention to adopt (measured as progress toward Stages 4-5).*

# 2. Methods

To answer our research questions and test the hypotheses above, we devised a Stages of Change measure and an experimental test of whether this measure could be used to determine whether the stage matters for the effectiveness of a specific security intervention: a one-page educational handout about 2FA. The resulting materials and our recruitment and experimental protocols were submitted for review by our Institutional Research Board and approved as exempt from human subjects regulation under U.S. 45 CFR 46. These are included in the Appendices and Supplemental Documents.

## 2.1 Creating a Stages of Change Measure

We identified several highly-cited research reports, literature reviews and meta-analyses that describe how the Stages of Change had previously been identified and/or measured. While some identified people's Stage of Change by coding interview responses [39,61] or administering a single survey item with a multiple-choice, single-response format [43,46,58], we chose to follow [40,47] in creating a staging algorithm that could be administered to a large-scale sample and that could help in screening out responses that indicate a participant was not paying sufficient attention to the questions to accurately classify their stage.

For this algorithm, we wrote five questions corresponding to Stages 1-5 (*Precontemplation* to *Maintenance*) with a Likert-type response set ranging from 1=*Strongly Disagree* to 5=*Strongly Agree*. Following the advice of [47], we made sure to clearly and explicitly define our target behavior of two-factor authentication in the instructions, and to state all the criteria needed for each stage: whether or not they perform the behavior currently, and how distant in time they might put this behavior into action (with benchmarks of 30 days and 6 months, per [44,47]). We underscored and bolded the key words in each item to make sure that participants would pay attention to them. See Appendix A for this tool.

To score the participant's Stage of Change, we first determined for which questions participants had marked *Agree* or *Strongly Agree*, and then assigned them to the highest Stage with which they agreed – for example, if they had marked Agree for both the statement corresponding to Stage 2 and for the statement corresponding to Stage 3, we marked them as being in Stage 3, in line with [40,56]. Participants who marked Agree or Strongly Agree for *contradictory* statements (such as for the Stage 1 and Stage 5 statements) were scored 0, and their data was excluded; likewise, participants who did not mark Agree or Strongly Agree for *any* of the five statements were scored 0, and their data also excluded. This algorithm would give us confidence that the results truly reflected the mindsets of participants and was not biased by inattentive answers.

## 2.2 Creating a Test of a Stage 1-Matched Intervention

We chose to use Amazon Mechanical Turk as the experimental platform for our study. This platform enabled us to easily prequalify participants, to deliver survey and experiment materials at arms' length, and to recruit a large sample for the research. We were able to deliver a link to the main experiment on Qualtrics as well as a short pretest and posttest Stages of Change measure coded directly in Mturk. Amazon also was a desirable site for this research because it offers Mturk workers the option – but not the requirement - to enable two-factor authentication under the parent site's Account Settings. This helped us to design an experiment that would have direct, real-world relevance for participants and that would benefit them apart from compensation by boosting their awareness of Amazon 2FA. See Table 1 for descriptive statistics about those who took the main survey.

*Table 1: Demographics and security breach experiences for N=142 participants who completed the pretest, main survey and posttest. This data was used to test H2 and H3.*

| N | 142 |
|---|---|
| **Age range** | |
| 18-29 | 28.9% |
| 30-39 | 42.3% |
| 40-49 | 17.6% |
| 50-59 | 7.7% |
| 60 or older | 3.5% |
| **Gender** | |
| Male | 38.0% |
| Female | 61.3% |
| Nonbinary or gender non-conforming | 0.7% |
| **Education** | |
| Some high school | 0.7% |
| High school degree or equivalent | 10.6% |
| Some college, assoc. or tech. degree | 36.6% |
| Bachelor's degree | 37.3% |
| Graduate or professional degree | 14.8% |
| **Yearly household income** | |
| Up to $25,000 | 18.3% |
| $25,000 to $49,999 | 33.1% |
| $50,000 to $74,999 | 19.0% |
| $75,000 to $99,999 | 12.7% |
| $100,000 or more | 16.9% |
| **How frequently or infrequently have you personally been the victim of a breach of security (e.g. hacking, malware)?** | |
| Very infrequently | 52.1% |
| Infrequently | 36.6% |
| Neither infrequently or frequently | 7.0% |
| Frequently | 4.2% |
| Very frequently | 0.0% |
| **How frequently or infrequently has someone close to you been the victim of a breach (e.g. hacking, malware)?** | |
| Very infrequently | 35.9% |
| Infrequently | 48.6% |
| Neither infrequently or frequently | 11.3% |
| Frequently | 3.5% |
| Very frequently | 0.7% |
| *How much have you heard/read in last year about breaches?* | |
| None at all | 1.4% |
| A little | 18.3% |
| A moderate amount | 43.7% |
| A lot | 32.4% |
| A great deal | 4.2% |

We designed an experimental test of this Stages of Change model to show *awareness-oriented information about 2FA* to participants who were prequalified by our staging algorithm into one of two buckets: Stage 1 of Security Behavior Change (*Precontemplation*), due to their agreement with a statement that they do not and *will not* enable Amazon 2FA for their account within the next six months; or Stage 2-3 of Security Behavior Change (*Contemplation-Preparation*), due to agreeing with a statement that they do not but *might* enable Amazon 2FA for their account, either within two to six months (Stage 2) or within the next 30 days (Stage 3). We decided to collapse Stage 2 and Stage 3 into one category for this test in order to facilitate a comparison with Stage 1 results [60], in line with prior work that had noted the difficulty of distinguishing the degree of behavior intention between Stage 2 and Stage 3 [23,44,47] or chosen to not include Stage 3 at all [46].

We then randomly assigned participants in these buckets to one of four experiment groups: exposure to a one-page handout that emphasizes the process of enabling 2FA (Group A); exposure to a one-page handout that emphasizes social norms of 2FA use (Group B); exposure to both a process-emphasis handout and a social-norms-emphasis handout (Group Both); and a control condition in which participants were shown a one-page handout of similar design about a non-2FA topic that is also relevant to Mturk study participants: the scientific method (Group Control). All 2FA handouts contained the definition of 2FA that was used in the staging algorithm (Appendix A). To further control for presentation and content effects, we created two versions of the process-emphasis and social-norms-emphasis handouts, one with text alone and one with text plus figures, and randomly served one or the other.

*Table 2: The 2x4 within-subjects experiment design allows for comparisons of stage progress and handout effectiveness.*

|  | 2FA process handout (Group A) | 2FA norms handout (Group B) | Both handouts (Group Both) | Non-2FA handout (Group Control) |
|---|---|---|---|---|
| **Stage 1** | A1 or A2 | B1 or B2 | A1B1 or B1A1 or A1B2 or B2A1 or A2B1 or B1A2 or A1B2 or B2A1 | Control |
| **Stages 2-3** | A1 or A2 | B1 or B2 | A1B1 or B1A1 or A1B2 or B2A1 or A2B1 or B1A2 or A1B2 or B2A1 | Control |

All handouts were created from a single predesigned template that we customized using the website Canva.com and then pilot-tested with n=6 colleagues, who suggested improvements for ease of reading and clarity of content. See Figure 2 for thumbnails of the handouts and Supplemental Documents for the full-size PDFs of the handouts used, and Appendix B for the flow and text of the main survey.

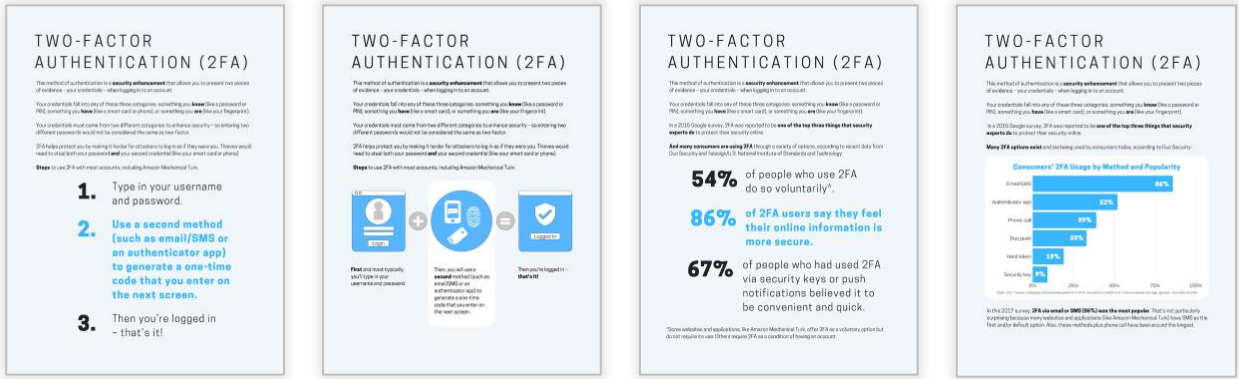

*Figure 2: The four 2FA handouts that we designed for the experiment (from left): A1 (process emphasis, text only), A2 (process emphasis, both text + figures used), B1 (social norms emphasis, text only), and B2 (social norms emphasis, both text + figures used). All handouts used the same 2FA definition as in Appendix A. See Supplemental Documents for full-size PDFs, including the control.*

After three days, we began again surveying participants as to their current Stage of Change for adopting Amazon 2FA, using the same algorithm that was used to qualify them for the study. The time frame of three days was chosen to give participants time to retain and process the information that they had acquired in the handout, to act on any inquisitiveness about 2FA specific to Mturk (Amazon's term is Two-Step Verification, although the authors found their FAQ online by using the search term "Amazon 2FA"), and/or to enable for their accounts. We then computed the difference between the pretest and posttest Stage of Change to split groups into those who had progressed toward Action and Maintenance (Stages 4-5), and those who had not.

# 2.3 Test Procedure

We prequalified Mturk workers using the staging algorithm described above, in a five-item survey coded directly into the website's interface to minimize the time needed to answer it. This initial Human Intelligence Task (HIT) was compensated at $0.02 and advertised as a

qualification test for a $1.50 "UX survey" to be posted within the following 72 hours. See Figure 3 for the distribution of pretest 2FA stages and Appendix A for the staging algorithm.

We then advertised our $1.50 "UX survey" to Mturk workers in Stage 1 and Stage 2-3 by using a custom qualification that hid the HIT from all but these specific workers, with 46.5% accepting it and submitting responses. The resulting sample of $n$=294 was used to test H1 by comparing the staging algorithm results with their SA-6 and SeBIS scores. See Appendix B for the text of the main survey.

Once three days had passed, we began advertising posttest HITs to again ask workers the five questions used in the staging algorithm in the pretest survey. We stopped collecting posttest survey responses once 10 days had passed, having reached a response rate of 48.3%. Once unreliable data had been excluded (see Results), this resulted in a sample of $n$=28 in Stage 1 and $n$=114 in Stages 2-3 from our pretest who took part in the main survey and the posttest survey and whose responses were deemed valid in all three surveys. This sample was used to test H2-3 by comparing the difference between their pretest and posttest scores with which handout they had been exposed to in the main survey.

# 3. Results

## 3.1 Validating a Stages of Change Measure

Figure 3, the histogram of the five-item staging algorithm's results, indicates that it measured sufficient variances in people's intention to adopt Amazon 2FA and that enough participants answered the questions in a way that made logical sense (77.9%). Of the *n*=1000 workers who took the test, 17.5% were identified as being in Stage 1 (*Precontemplation*) and 45.7% were identified as being in Stages 2-3 (*Contemplation* and *Preparation*).

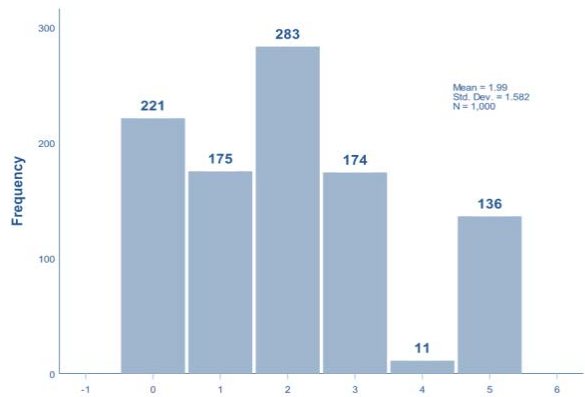

*Figure 3: Pretest distribution of Stages of Change for Amazon 2FA (n=1000). Stage 0 means there was no valid indication of 2FA stage; Stage 1 indicates no intention to adopt Amazon 2FA; Stage 2 indicates openness to adopting Amazon 2FA in the next 2-6 months; Stage 3 indicates an intention to adopt Amazon 2FA in the next 30 days; Stage 4 indicates recent adoption of Amazon 2FA (within 30 days of the survey); and Stage 5 indicates long-term adoption of Amazon 2FA (more than 30 days from the survey).*

Moreover, we found that a low Stage of Change (1) was associated with lower scores on two previously validated security scales, the six-item Security Attitude scale (SA-6) [31] and the 16-item Security Behavior Intentions Scale (SeBIS) [27], while the higher Stages of Change (2-3) were associated with higher scores on these scales. An independent-samples Kruskal-Wallis test found these differences to be statistically significant for both SA-6 (Stage 1 [*M*=3.26, *SD*=.80] vs. Stage 2-3 [*M*=3.47, *SD*=.70], *p*<.05) and SeBIS (Stage 1 [*M*=3.44, *SD*=.67] vs. Stage 2-3 [*M*=3.70, *SD*=.58], *p*<.005). We used a non-parametric means test because the stages are not normally distributed.

- **H1a** *Supported.*
- **H1b** *Supported.*

These results demonstrate that a staging algorithm can be validly used in place of or alongside these scales to screen participants for a specific Stage of Change and to quantify and compare their progress toward Stages 4-5 (*Action and Maintenance*).

## 3.2 Evaluating the 2FA Informational Handouts

We first analyzed the data for the time that participants spent on each handout page of the survey and for answers to control questions that we asked all participants: a four-item usability rating adapted from [6], and a two-item test of their recall of information shown in the handouts. Using a stem- and-leaf plot and a boxplot of the data distributions, we identified and excluded one participant who spent an unusually long time on a handout (more than 250 seconds). We also excluded several who rated an assigned handout's usability as below 3.00 on a 1.00-5.00 scale, and several who answered 0/2 recall questions correctly (*n*=21 total responses removed) as unlikely to have retained any information.

The trimmed dataset showed that the handouts had roughly equal usability scores, that participants had slightly more trouble recalling facts from the process-emphasis (A) handouts than from the social-norms-emphasis (B) handouts, and that participants viewed handouts A1 and B2 for at least 10 seconds longer than A2 and B1. Using an independent samples Kruskal-Wallis test, we found that the total time spent viewing handouts did not make a difference in progress toward 2FA adoption (Progress [$M$=66.8, $SD$=50.2] vs. No Progress [$M$=67.4, $SD$=40.0], $p$=n.s.).  See Table 3 for timing, usability and recall statistics by handout.

*Table 3: For the dataset used for answering RQ2, a comparison of time spent viewing the handouts, the mean UX score, and the percentage who answered both recall questions correctly. Participants shown handout A1 spent the most time on average viewing it and struggled most with recall questions. Not all participants who viewed handouts also took part in the posttest.*

| Handout | $n$ | Time spent (s) | | UX score | | 2/2 recall score |
|---------|-----|------|-----|------|-----|------|
| | | $M$ | $SE$ | $M$ | $SE$ | |
| Control | 38 | 46.9 | 9.4 | 4.2 | 0.1 | 97.4% |
| A1 | 87 | 62.2 | 13.3 | 4.3 | 0.1 | 81.6% |
| A2 | 102 | 43.8 | 3.8 | 4.2 | 0.1 | 84.3% |
| B1 | 100 | 50.4 | 4.7 | 4.2 | 0.1 | 92.0% |
| B2 | 85 | 60.1 | 10.6 | 4.3 | 0.1 | 91.8% |

For H2, our analysis of this trimmed dataset shows that the 2FA handout with the social norms emphasis was not as effective as the handout that emphasizes the process of adopting 2FA in moving participants toward 2FA adoption. For H3, the combination of the two handouts was significantly more effective than the social norms handout or the control. The difference in progress among the four groups was significant in a chi-square test, which is an appropriate means test for smaller samples ($x^2$[3,142]=7.95, $p$<.05).

- **H2** *Partially supported.*
- **H3** *Supported.*

See Figure 4 for a chart of how many participants showed progress toward later Stages of Security Behavior Change for a 2FA process handout (A) vs. a handout oriented toward social norms of 2FA use (B), viewing both types of handouts (Both), or viewing the non-2FA handout (Control).

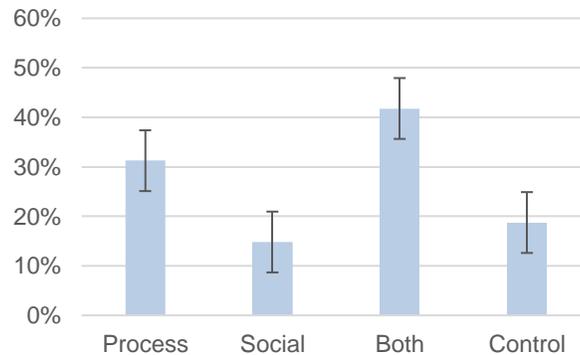

*Figure 4: A comparison of how many in each experimental group who also completed the posttest survey showed progress toward Stage 4-5. Those exposed to a process-emphasis handout (in Group A or Group Both) did better than those who saw only a social-norms handout or control.*

## 3.3 Evaluating Stage Matching

Additionally, we found a statistically significant difference in the number of Stage 1 participants who progressed toward later Stages of Change (n=14, 57.0%), compared with the progress of Stage 2-3 participants toward later stages (n=28, 27.7%) who were exposed to awareness-oriented 2FA handouts ($x^2$[1,126]=7.211, *p*<.01).

These results support the hypothesis that awareness-oriented information about 2FA would be a better match for those in Stage 1 vs. Stages 2-3, as predicted by prior work; a majority of Stage 1 made progress toward Stages 4-5 regardless of 2FA group vs. those in Stages 2-3, though the inter-stage difference for Stage 1 posited in H2a was not significant, while for H2b, it was statistically significant.

- ***H4a*** *Supported.*
- ***H4b*** *Supported*.

To test for whether a combination of these handouts might show a difference in effectiveness, we randomly exposed subgroups of Stage 1 and Stages 2-3 to one of both types of handouts, with either the process-emphasis handout first (AB) or the social-norms handout first (BA). Again, we saw a statistically significant difference in the number of Stage 1 participants who progressed toward later Stages of Security Behavior change (n=10, 76.9%) compared with the progress of Stage 2-3 participants (n=18, 33.3%) who were exposed to a combination of handouts ($x^2$[1,67]=8.184, $p<$.005).

- **H4c Supported.**

See Figure 5 for a comparison of progress toward Stages 4-5 of 2FA adoption by pretreatment stage.

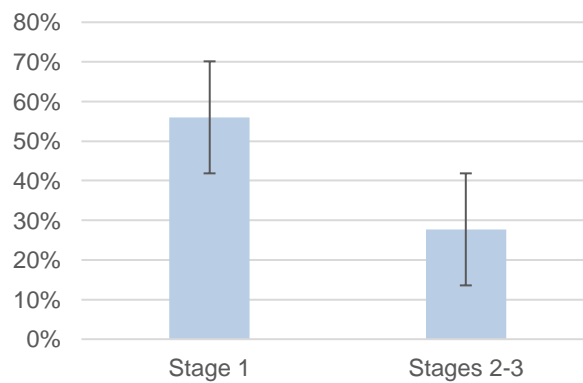

*Figure 5: A majority in the Stage 1 group (n=14, 57.0%) who viewed 2FA handouts showed progress toward Stage 4-5, while only a minority in the Stage 2-3 group (n=28, 27.7%) showed progress toward Stages 4-5 ($x^2$[1,126]=7.211, $p<$.01).*

# 4. Discussion

Our methods and data analysis show that our TTM-adapted staging algorithm is valid and effective in sorting participants according to their degree of intention to adopt expert-recommended security measures such as 2FA. Our experiment found that using a stage-matched handout intervention led to a practical difference in progress toward adopting 2FA for those in Stage 1 versus those in Stage 2-3, as predicted by the TTM, and that the awareness-oriented information failing to move a significant portion of Stage 2-3 participants toward Stages 4-5.

*Table 4: A summary of hypotheses tested and whether the results support them.*

| **_Hypothesis Tested in This Study_** | **_Supported?_** |
|---|---|
| **H1a:** *Stronger intention to adopt a security behavior is positively associated with a higher security behavior change score.* | Yes |
| **H1b:** *Security attitude is positively associated with a higher security behavior change score.* | Yes |
| **H2:** *Exposure to information emphasizing either social norms or process of 2FA use will increase the likelihood of progressing toward intention to adopt.* | Partially |
| **H3:** *Exposure to information emphasizing both social norms and process of 2FA use will increase the likelihood of progressing toward intention to adopt.* | Yes |
| **H4a:** *For people in precontemplation (stage 1), exposure to information about 2FA will increase progress toward intention to adopt (measured as progress toward Stages 4-5).* | Yes |
| **H4b:** *For people in contemplation or preparation (stages 2 or 3), there will be no influence of exposure to a 2FA handout.* | Yes |
| **H4c:** *For people in precontemplation (stage 1), exposure to a combination of 2FA information will increase progress toward intention to adopt (measured as progress toward Stages 4-5).* | Yes |

## 4.1 Advantages of the Stages of Security Change Measure

Our tests of validity show that our staging algorithm is significantly positively associated with existing measures of security attitude and security behavior intention. This supports its use in place of these scales for measuring security attitude and security behavior intention. Our measure also showed evidence of weeding out participants who were not reading the items and responses, while also validly measuring responses of participants who did read the text. This shows the measure's utility for researchers who are conducting online research and who need simple ways to verify that they are receiving good-faith survey responses.

Our staging algorithm goes one step further than existing attitude and intention scales, however, by incorporating a behavioral measure – whether someone has, in fact, enabled 2FA authentication (has not enabled = Stages 1-3, has enabled = Stages 4-5). Only one recent study has attempted to create a reusable measure of users' self-report security behavior to be used alongside measures of attitude and behavior intention [31]. Our staging algorithm thus has utility for researchers as a practical measurement of attitude, intention and behavior in one simple, five-item survey.

## 4.2 Advantages of the Stages of Security Change Model

Using criteria laid out by Prochaska et al. [45], we find the Stages of Security Change Model [29,30,55] to be a desirable framework for creating and assessing usability interventions.

### 4.2.1 Clarity

We find that the conceptual model as laid out by Faklaris [29,30] and Ting [55] defines its propositions and concepts in a way that is easily understandable and that our study demonstrated has content and construct validity.

### 4.2.2 Consistency

We find acceptable fit between the Stage 1 and Stages 2-3 categories and the use of an awareness intervention for nudging Stage 1 users to a later Stage of Change.

### 4.2.3 Parsimony

The model as laid out in Figure 1 is uncomplicated and explains and predicts behavior in a simple manner.

### 4.2.4 Testable

We were able to create a method for reliably and validly testing a person's Stage of Security Behavior Change.

### 4.2.5 Empirical Adequacy

The claims made in the Stage of Security Behavior Change model are congruent with evidence from our study.

### 4.2.6 Productivity

The model adds to the knowledge base in usable security, builds on prior work and generates ideas for future studies.

### 4.2.7 Utility

The model is useful for differentiating which users are most primed for an awareness- or action-oriented intervention.

### 4.2.8 Practicality

Our theory-motivated awareness handouts produced progress toward 2FA adoption (Stages 4-5).

## 4.3. Limitations and Future Work

Our use of convenience sampling on Amazon Mechanical Turk limits the generalizability of our findings. The response rate and size of our Stage 1 participant bucket may be evidence of a self-selection bias that influenced our results. We also did not conduct follow-up interviews with survey participants, which limits our ability to identify how and why we found significant effects for the Stage 1 bucket of participants versus the Stage 2-3 bucket.

While all handouts' usability was rated at least a 4.00 on the 1.00-5.00 scale, some intriguing differences emerged on recall statistics (Table 3). Nine in 10 participants who viewed the B1 and B2 handouts were able to recall the emphasized social information, while only about eight in 10 participants could recall the emphasized process information about 2FA. Yet it was the group that only viewed the social-norms handouts that showed the least progress toward adopting 2FA vs. the control group. It may be that, regardless of usability of the information or the ease of immediate recall, the design strategy of juxtaposing social-norms information with other practical information about 2FA on a single page was distracting enough to cancel out a benefit from being exposed to social information about 2FA. We will test a new operationalization of social norms in a future study.

The finding that the group exposed to two different handouts did so much better than those shown a single handout may also indicate that repetition of 2FA facts in the content of the recall questions (no feedback was given as to correct answers) may have contributed to Group A's better progress than Group B toward 2FA adoption, regardless of their relatively poorer recall of the highlighted 2FA information. We will test whether a combination of information can lift progress for all conditions in a future study.

Finally, our choice to focus on raising 2FA awareness via handout materials leaves open the question of which specific 2FA interventions are a better match for Stages 2-3, or whether information delivered through other means such as video tutorials [6] or in populations outside of Mturk [14] are effective with the Stages of Change model. We plan to explore these alternatives in a future study.

# 5. Conclusion

Our study makes two contributions to usability and usable security. First, we have adapted and validated a method for measuring end users' Stages of Change progress toward voluntarily deciding to enable security measures such as 2FA. Second, we have found empirical support for using the Stages of Security Behavior Change model to target awareness-oriented interventions to people in Stage 1 (no intention to voluntarily adopt recommended measures such as 2FA) but not to those in Stages 2-3 (some degree of intention to adopt recommended measures such as 2FA). We hope that these results, and the examples of effective 2FA communication in our handouts, will help industry practitioners and researchers alike to more efficiently train end users to comply with expert recommendations for their account security and to encourage voluntary adoption.

Our study provides researchers and industry professionals with valuable tools beyond system log data with which to measure computer users' acceptance and resistance to complying with security policies and following security advice. We hope this will help the fields of psychology, human-computer interaction and cybersecurity to explain, predict and influence the behaviors of computer users in ways that head off social engineering attacks and improve the security of computer networks.

# Acknowledgements

This work was generously supported by the U.S. National Science Foundation, grant no. CNS-1704087. The first author also is grateful for fellowship support from the CyLab Security and Privacy Institute and the Center for Informed Democracy and Social Cybersecurity, both at Carnegie Mellon University, and for feedback on previous versions of this work by reviewers for the USENIX Symposium on Usable Privacy and Security (SOUPS), by the Workshop on Security Information Workers (WSIW), by members of the Connected Experiences Lab at the HCII, and many others. Sponsors were not involved in any phase of research or article preparation.

# Appendix A: Staging algorithm

*[Definition given before the five statements:]*

**Two-factor authentication**, or **2FA**, is a security enhancement that allows you to present two pieces of evidence – your credentials – when logging in to an account.

Your credentials fall into any of these three categories: something you know (like a password or PIN), something you have (like a smart card), or something you are (like your fingerprint).

Amazon gives its users the option to enable 2FA for their accounts at the "Login & security" page under "Advanced Security Settings." (Note: The researchers have no affiliation with Amazon other than as customers and requesters and are not conducting this survey or sharing project data with Amazon.)

***On a scale of 1 to 5 ("*Strongly disagree*" to "*Strongly agree*"), rate your agreement with each statement.***

1. I **do not** and **will not** use 2FA with my Amazon account **in the next two to six months**.

<div align="center">

*Strongly disagree*   1   2   3   4   5   *Strongly agree*

</div>

2. I **do not** but **might** use 2FA with my Amazon account **in the next two to six months**.

<div align="center">

*Strongly disagree*   1   2   3   4   5   *Strongly agree*

</div>

3. I **do not** but **will** use 2FA on my Amazon account **in the next 30 days**.

<div align="center">

*Strongly disagree*   1   2   3   4   5   *Strongly agree*

</div>

4. I **do** use 2FA on my Amazon account **as of today** and started using it **no more than 30 days ago**.

*Strongly disagree*   1  2  3  4  5  *Strongly agree*

5. I **do** use 2FA on my Amazon account **as of today** and have **within the past two to six months**.

*Strongly disagree*   1  2  3  4  5  *Strongly agree*

# Appendix B: Main survey

## B.1 Survey Flow After Participant Consent

*Random assignment to one of nine conditions, then further randomization within groups to four*

*additional conditions:*

1. Condition: Control
2. Condition: Treatment A1
3. Condition: Treatment A2
4. Condition: Treatment B1
5. Condition: Treatment B2
6. Condition: Group A1B1
   a. Treatment A1B1
   b. Treatment B1A1
7. Condition: Group A1B2
   a. Treatment A1B2
   b. Treatment B2A1
8. Condition: Group A2B1
   a. Treatment A2B1
   b. Treatment B1A2
9. Condition: Group A2B2
   a. Treatment A2B2
   b. Treatment B2A2

- Security Attitude (6 Questions [31])
- Security Behavior Intention (16 Questions [27])
- Amazon 2FA (8 Questions, to check that the participant values their Amazon account and to classify their Stage of Security Behavior Change for Amazon 2FA [Appendix A])
- General Questions about Information, Security and Privacy (4 Questions, to check that they are paying attention and to assess recent security breach exposures, adapted from [31])
- Demographics (6 Questions, adapted from [31])

## B.2 Evaluation Questions for Each Handout

[*Instructions*:]

Please look over the infographic below. We will then ask you several questions to evaluate the design.

*[Collect time in seconds until Page Submit fires]*

*---- Next page*

*[Recall Questions for Control:]*

Amazon Mechanical Turk is used to help perform experiments.

o      True [correct]

o      False

Analyzing data is **not** an important part of the scientific method.

o      True

o      False [correct]

*[Recall Questions for A1 and A2:]*

In 2FA, the "second factor" often is something you have, such as phone, card or token.

o      True [correct]

o      False

You can use the second factor to generate a one-time code that you enter on screen.

o      True [correct]

o      False

*[Recall Questions for B1:]*

Most recently surveyed 2FA users say they feel that their information is more secure.

o      True [correct]

o      False

Only 10% of those surveyed say they are using 2FA voluntarily.

o    True

o    False [correct]

*[Recall Questions for B2:]*

Only one method exists for using 2FA.

o    True

o    False [correct]

The most popular method for using 2FA is a hard token.

o    True

o    False [correct]

*---- Next page*

*[Instructions:]*

Please rate your level of agreement with the following statements.

*Matrix table scoring columns: 1=Strongly Disagree, 2=Somewhat Disagree, 3=Neither*

*Disagree nor Agree, 4=Somewhat Agree, 5=Strongly Agree:*

I found this infographic to be **easy to understand**.

I found this infographic to be **interesting**.

I found this infographic to be **informative**.

I found this infographic to be **useful**.

## B.3 Questions to Assess Value of Amazon Accounts

*Scale for answering questions: 1=Strongly Disagree, 2=Somewhat Disagree, 3=Neither Disagree nor Agree, 4=Somewhat Agree, 5=Strongly Agree:*

My Amazon account is **important to me**.

My Amazon account **holds data that I want to protect**.

# Supplemental Materials

Full-sized handouts that participants in each treatment were directed to view are available for download at [https://corifaklaris.com/files/2FA_Handouts.zip](https://corifaklaris.com/files/2FA_Handouts.zip).